\documentclass[twocolumn,amsmath,amsfonts,aps,prc,floatfix]{revtex4}
 \usepackage{graphicx} 
 
\begin{document} 
 
\title{Charged particle's $p_T$ spectra and elliptic flow in $\sqrt{s_{NN}}$=200 GeV Au+Au collisions: QGP vs. hadronic resonance gas}
  
\date{\today}
  
\author{A. K. Chaudhuri} 
\email{akc@veccal.ernet.in} 
\author{Victor Roy}
\email{victor@veccal.ernet.in}
\affiliation{Variable Energy Cyclotron Centre, 1-AF, Bidhan Nagar,
Kolkata - 700 064, India} 
 \begin{abstract}
  
  We show that if the hadronic resonance gas (HRG), with viscosity to entropy ratio $\eta/s\approx$0.24, is physical at  temperature $T\approx$220 MeV, 
charged particles $p_T$ spectra and elliptic flow in Au+Au collisions at RHIC, over a wide range of collision centrality
do not distinguish between initial QGP fluid and initial hadronic resonance gas. 
Unambiguous identification of bulk of the matter produced in Au+Au collisions require clear demonstration that HRG is unphysical at temperature $T>$200 MeV.
 It calls for precise lattice simulations with realistic boundary conditions. 
\end{abstract}
\maketitle  

 
Experiments at Relativistic Heavy Ion Collider (RHIC) produced convincing evidences that in central and mid central Au+Au collisions, a hot, dense, strongly interacting matter is created\cite{BRAHMSwhitepaper},\cite{PHOBOSwhitepaper},\cite{PHENIXwhitepaper},\cite{STARwhitepaper}. Whether the matter can be identified as the lattice QCD 
predicted Quark Gluon Plasma(QGP) or not is still a question of debate. The problem is closely related to quark confinement; quarks are unobservable and any information about the initial state has to be obtained from observed hadrons only. QGP, even if produced in Au+Au collisions, is a transient state,  it expands, cools, hadronises, cools further till interactions between the hadrons become too weak to continue the evolution. Hadronisation is a non-perturbative process. 
Whether or not the hadronisation process erases any memory of the constituent quarks is uncertain. If the hadronisation process erases the memory, from the observed hadrons one can not comment on the initial QGP phase. Present search for QGP at RHIC is on the premise that the hadronisation process does not erase the memory and from the observed hadrons, using a dynamical model like hydrodynamics, one can back-trace to  the initial QGP phase. Hydrodynamic equations are closed with an Equation of State (EOS) and one can incorporate the possibility of phase transition in the model. Indeed, a host of experimental data in Au+Au collisions at RHIC is well explained in a hydrodynamical model with QGP in the initial state \cite{QGP3}. The alternative, namely hadronic resonance gas (HRG) in the initial state
 generally give poorer description to the data \cite{Kolb:2000fha}, that too with  HRG at very high temperature, $T_i\approx$ 270 MeV.  
At such high temperature hadrons density is very large  $\rho_{had} \sim$ 4 $fm^{-3}$ and it is difficult to believe that they retain their individual identity. However, situation changes if viscous effects are included. 
Model calculations \cite{Gorenstein:2007mw},\cite{Demir:2008tr},\cite{Muronga:2003tb},\cite{Pal:2010es}   indicate that for a hadronic resonance gas, viscosity to entropy ratio is considerably larger than the ADS/CFT limit, $\eta/s \geq 1/4\pi$  \cite{Policastro:2001yc}. Since entropy is generated during evolution, unlike an 'ideal' HRG, a 'viscous' HRG can be initialized at a lower temperature such that hadrons retain their identity yet reproduce the experimental multiplicity. In the following, we show that  if the hadrons retain their identity at temperature $T\approx$220 MeV, charged particles $p_T$ spectra and elliptic flow in 0-40\% Au+Au collisions, do not distinguishes between initial HRG with viscosity to entropy ratio $\eta/s \approx$0.24 and initial QGP with $\eta/s$=0.08. Unambiguous identification of the matter with QGP will require clear proof that HRG has a limiting temperature $T\leq$ 200 MeV or the confinement-deconfinement transition occur at a temperature $\leq$ 200 MeV. 

Most reliable information about the confinement-deconfinement transition temperature is obtained in lattice simulations of QCD. 
It is now established that the confinement-deconfinement transition is not a thermodynamic phase transition, rather a cross-over. Since it is cross-over, there is no unambiguous temperature where the transition take place.   
Inflection point of the Polyakov loop is generally quoted as the (pseudo)critical temperature for the confinement-deconfinement transition. Currently, there is debate over the value of the pseudo critical temperature.   HotQCD collaboration, with physical strange quark mass and somewhat larger than physical u and d quark masses ($m_s/m_{u,d}=10$) claimed that both the chiral and confinement-deconfinement transition takes place at a common temperature $T_c=192(4)(7)$. \cite{Cheng:2006qk},\cite{Cheng:2007jq},\cite{Detar:2007as},\cite{Karsch:2007dt},\cite{Karsch:2008fe}. 
For more physical u and d masses ($m_s/m_{u,d}$=20), the critical temperature shift by $\sim$ 5 MeV to lower  side \cite{Cheng:2009zi}. Wuppertal-Budapest collaboration, with physical strange and light quark masses ($m_s/m_{u,s}=28$), on the other obtained a different result. \cite{Aoki:2006br},\cite{Aoki:2009sc}, \cite{Fodor:2010zz},  \cite{Borsanyi:2010bp}.  Chiral transition temperature  $T_c$=157 (3)(3) MeV is $\sim$ 20 MeV less than the confinement-deconfinement transition temperature $T_c$=170(4)(4) MeV. Several other observables, e.g. quark susceptibility, pressure etc in Wuppertal-Budapest simulations is also shifted by 20-30 MeV compared to HotQCD.   The $\sim$ 20 MeV shift in the confinement-deconfinement transition temperature is not properly understood. Possible reasons could be larger physical quark masses in HotQCD simulations.
Also, one notes that though both the collaborations use staggered fermions, the fermion actions are different. It is still unproven that  in the continuum limit, staggered fermions represent QCD. The difference in the pseudo critical temperature may also be due to different 'staggered' fermion action in the two simulations. 
Additionally, lattice results for the transition  temperature can be up by $\sim$ 30 MeV. All the lattice simulation uses periodic boundary condition, while in a realistic situation e.g. heavy ion collisions, the deconfined region is bordered  by a confined phase. Exploratory quenched study suggests that with realistic boundary condition transition temperature can be up by 30 MeV \cite{Bazavov:2007zz}. 
Considering the uncertainties associated with lattice simulations, the pseudo critical temperature for the confinement-deconfinement transition temperature could be as high as $T_c\approx$ 220 MeV. 
Qualitatively, for hadron size $\approx$ 0.5 fm, limiting hadron density (such that hadrons are not overlapped) is $\rho^{had}_{limit}=1/V \approx$ 2 $fm^{-3}$. For HRG, $\rho_{had}\approx$ 2$fm^{-3}$ corresponds to limiting temperature $T_{limit}$= 220 MeV, value close to the highest possible pseudo critical temperature, as argued above. $T\approx$220 MeV is possibly the highest temperature at which hadrons can retain their individual identity. At higher temperature hadrons will overlap extensively and lose their identity.  

In the following we simulate Au+Au collisions in two scenarios, (i) no phase transition (NPT) scenario when a HRG with limiting temperature $T$=220 MeV is produced in the initial collisions and (ii) a phase transition (PT) scenario when QGP is produced. Space-time evolution of HRG/QGP fluid is obtained by solving,

\begin{eqnarray}  
\partial_\mu T^{\mu\nu} & = & 0,  \label{eq1} \\
D\pi^{\mu\nu} & = & -\frac{1}{\tau_\pi} (\pi^{\mu\nu}-2\eta \nabla^{<\mu} u^{\nu>}) \nonumber \\
&-&[u^\mu\pi^{\nu\lambda}+u^\nu\pi^{\nu\lambda}]Du_\lambda. \label{eq2}
\end{eqnarray}

Eq.\ref{eq1} is the conservation equation for the energy-momentum tensor, $T^{\mu\nu}=(\varepsilon+p)u^\mu u^\nu - pg^{\mu\nu}+\pi^{\mu\nu}$, 
$\varepsilon$, $p$ and $u$ being the energy density, pressure and fluid velocity respectively. $\pi^{\mu\nu}$ is the shear stress tensor (we have neglected bulk viscosity and heat conduction). Eq.\ref{eq2} is the relaxation equation for the shear stress tensor $\pi^{\mu\nu}$.   
In Eq.\ref{eq2}, $D=u^\mu \partial_\mu$ is the convective time derivative, $\nabla^{<\mu} u^{\nu>}= \frac{1}{2}(\nabla^\mu u^\nu + \nabla^\nu u^\mu)-\frac{1}{3}  
(\partial . u) (g^{\mu\nu}-u^\mu u^\nu)$ is a symmetric traceless tensor. $\eta$ is the shear viscosity and $\tau_\pi$ is the relaxation time.  It may be mentioned that in a conformally symmetric fluid relaxation equation can contain additional terms  \cite{Song:2008si}. Assuming longitudinal boost-invariance, 
the equations are solved  with the code 'AZHYDRO-KOLKATA'
\cite{Chaudhuri:2006jd},\cite{Chaudhuri:2008ed},\cite{Chaudhuri:2008sj},\cite{Chaudhuri:2008je} in ($\tau=\sqrt{t^{2}-z^{2}},x,y,\eta=\frac{1}{2}ln\frac{t+z}{t-z}$) coordinates.
  
Hydrodynamic equations (Eq.\ref{eq1} and \ref{eq2}) are closed only with an equation of state (EOS) $p=p(\varepsilon)$. In the NPT scenario, we use EOS for the non-interacting hadronic resonance gas, comprising all the resonances with mass $m_{res} \leq $2.5 GeV. In the PT scenario, we use a  lattice based EOS \cite{Chaudhuri:2009uk}, where the high temperature phase is modeled by the recent lattice simulation \cite{Cheng:2007jq}.
At the cross-over temperature $T_{co}$=196 MeV, the EOS smoothly changes to that of the hadronic resonance gas.  
  
Solution of Eq.\ref{eq1} and \ref{eq2} require initial energy density, velocity
distribution in the transverse plane at the initial time. A freeze-out prescription is also needed. In the PT scenario, we assume that QGP fluid is thermalised in the time scale $\tau_i$=0.6 fm \cite{QGP3}. Compared to QGP, a hadronic resonance gas is expected to thermalise late. We assume the canonical value, $\tau_i$=1 fm as the initial time in the NPT scenario.
Both in the PT and NPT scenario, initial fluid velocity is assumed to be zero.
In an impact parameter ${\bf b}$ collision, 
the initial energy density is assumed to be distributed as 
 \cite{QGP3},

\begin{equation} \label{eq3}
\varepsilon({\bf b},x,y) = \varepsilon_0[(1-f)N_{part} ({\bf b},x,y)+ f N_{coll}({\bf b},x,y)]
\end{equation}  

  \noindent where $N_{part}({\bf b},x,y)$ and $N_{coll}({\bf b},x,y)$ are the transverse profile of participant numbers and binary collision numbers. $f$ 
in Eq.\ref{eq3} is the fraction of hard scattering. 
Most of the hydrodynamic simulations are performed with hard scattering fraction $f$=0.25 or  0.13 \cite{QGP3},\cite{Hirano:2009ah}. We assumed $f=0.25$  in PT scenario. Hard scattering fraction is assumed to be zero in the NPT scenario. $\varepsilon_0$ in Eq.\ref{eq3} is the central energy density of the fluid in   impact parameter ${\bf b}$=0 collision. In the PT scenario, for a fixed freeze-out temperature $T_F$=150 MeV, $\varepsilon_0$ is varied to best reproduce  experimental charged particles $p_T$ spectra in 0-10\% Au+Au collision. In the NPT scenario, we used the limiting value, $\varepsilon_0$=5.1 $GeV/fm^3$, corresponding to central temperature $T_i$=220 MeV. HRG initialised at central energy density $\varepsilon_0$=5.1 $GeV/fm^3$ do not produce adequate number of hadrons from the freeze-out surface at $T_F$=150 MeV.  We lower the freeze-out temperature to $T_F$=110 MeV. Dissipative hydrodynamics also require initialisation of the shear stress tensors $\pi^{\mu\nu}$. In both the scenarios, we initialise $\pi^{\mu\nu}$  at the boost-invariant values, $\pi^{xx}=\pi^{yy}=2\eta(x,y)/3\tau_i$, $\pi^{xy}=0$ \cite{Chaudhuri:2008je}.
For the relaxation time $\tau_\pi$, we use the Boltzmann estimate, $\tau_\pi=6\eta/4p$. 

 \begin{figure}[t]
\center
 \resizebox{0.35\textwidth}{!}{%
  \includegraphics{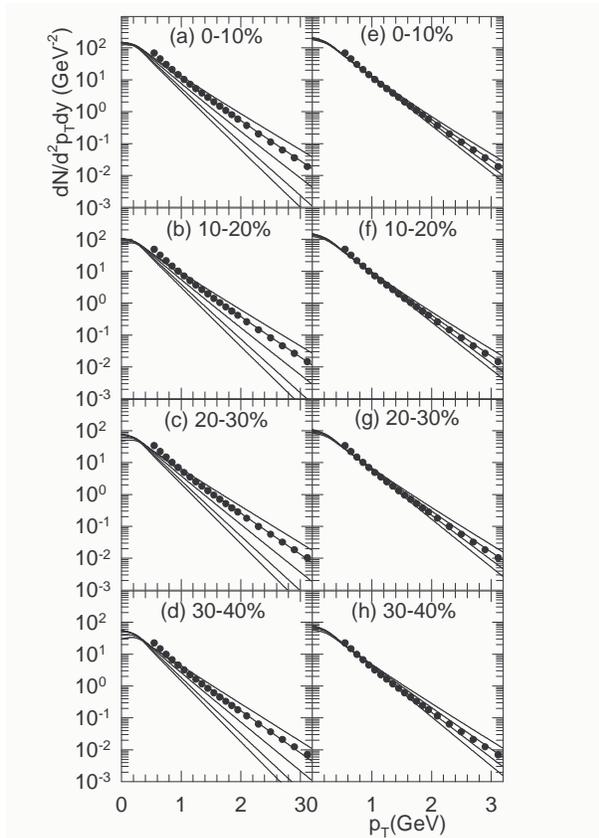}
}
\caption{PHENIX data \cite{Adler:2003au} for charged particles $p_T$ spectra   in  0-40\% Au+Au collisions are compared with 
hydrodynamic simulations in NPT (the left panel) and PT scenario (the right panel). See the text for details.}
  \label{F1}
\end{figure}

\begin{figure}[t]
\center
 \resizebox{0.35\textwidth}{!}{%
  \includegraphics{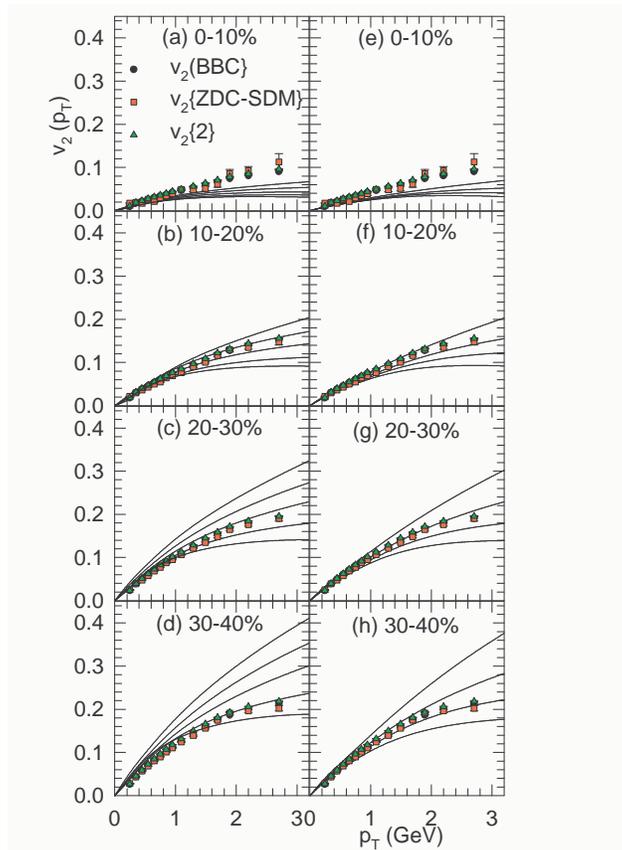}
}
\caption{(color online) same as in Fig.\ref{F1} but for elliptic flow.}
 \label{F2}
\end{figure}

Results of our simulations for charged particles $p_T$ spectra and elliptic flow in 0-40\% Au+Au collisions are shown in Fig.\ref{F1} and \ref{F2}. We have assumed that through out the evolution,  viscosity to entropy ratio remains a constant. In PT scenario, we have simulated Au+Au collisions for
four values of viscosity, (i)$\eta/s$=0 (ideal fluid) (ii)$\eta/s=1/4\pi \approx$ 0.08 (ADS/CFT limit), (iii)$\eta/s$=0.12  and (iv)$\eta/s$=0.16. 
Corresponding central energy densities are 
 $\varepsilon_0$=35.5, 29.1, 25.6 and 20.8 ($GeV/fm^3$) respectively. In the NPT scenario, we simulate Au+Au collisions for five values of viscosity, $\eta/s$=0, 0.08, 0.12, 0.24 and 0.30.
  
In Fig.\ref{F1},  we have compared simulated 
charged particles $p_T$ spectra in 0-40\% Au+Au collisions with the PHENIX data \cite{Adler:2003au}. 
The left panels (a-d) show the fit to the data in the NPT scenario.
The solid lines from bottom to top corresponds to $\eta/s$=0, 0.08, 0.16, 0.24 and 0.30 respectively. Simulation results in the PT scenario are shown in the right panel (e-h). The lines from bottom to top are for viscosity to entropy ratio $\eta/s$=0, 0.08, 0.12 and 0.16 respectively. 
One observes that ideal HRG, initialised to central temperature $T_i$=220 MeV, do not explain the $p_T$ spectra, data are largely under predicted. Discrepancy between simulated spectra and experiment diminishes with increasing viscosity and data are best explained with $\eta/s$=0.24. Description to the data  deteriorates if viscosity is further increased.
In the PT scenario also, best fit to the 0-40\% data is obtained in viscous QGP evolution, with $\eta/s$=0.08-0.12. Ideal QGP or QGP fluid with $\eta/s>0.12$ give comparatively poorer description. To be quantative about the fit to the data in two different scenarios,  we have computed $\chi^2$ values for the fits obtained to the 0-40\% data,  
  
\begin{equation}
\chi^2/N=\frac{1}{N}\sum_i \frac{(ex(i)-th(i))^2}{err(i)^2}
\end{equation}

\begin{figure}[t]
\center
 \resizebox{0.40\textwidth}{!}{%
  \includegraphics{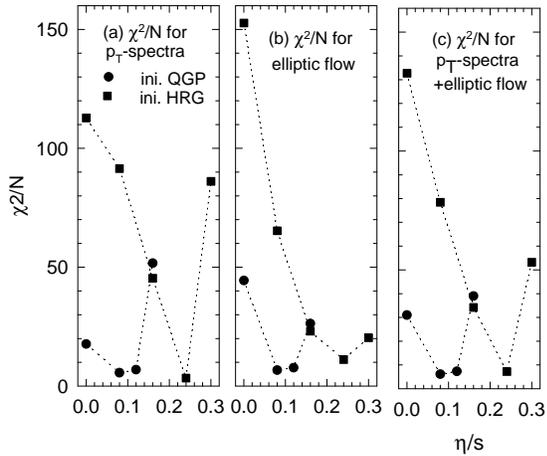}
}
\caption{The filled circles and squares are $\chi^2/N$ for (a)$p_T$ spectra, (b) elliptic flow and (c) $p_T$ spectra+elliptic flow, in 0-40\% Au+Au collisions, in
PT (initial QGP) and NPT (initial HRG) scenarios.}  \label{F3}
\end{figure}

We include both the statistical and systematic errors in the analysis.
The $\chi^2/N$ values are shown in Fig.\ref{F3}a. In the NPT scenario, $\chi^2/N$ as a function of $\eta/s$ shows a minima at $\eta/s$=0.24,  when $(\chi^2/N)_{min}$=3.3. It is very interesting to note that the $\eta/s$=0.24, obtained from the analysis is in close agreement with theoretical estimates of viscosity to entropy ratio of a hot hadronic resonance gas, $\eta/s$=0.24-0.30   \cite{Gorenstein:2007mw}.
In the PT scenario also, best fit to the 0-40\% data is obtained in viscous QGP evolution, with $\eta/s$=0.08, when $(\chi^2/N)_{min}$=5.6. Comparable fit  is obtained with QGP viscosity $\eta/s$=0.12. Quantatively,  with the initial conditions as used here, charged particles $p_T$ spectra in 0-40\% Au+Au collisions are better explained with HRG than QGP in the initial state.

In fig.\ref{F2}, we have compared simulated elliptic flow in 0-40\% Au+Au collisions with PHENIX measurements \cite{Afanasiev:2009wq}.  PHENIX collaboration obtained $v_2$ from two independent analysis,
(i) event plane method from two independent subdetectors, $v_2\{BBC\}$ and 
$v_2 \{ZDC-SMD\}$ and (ii) two particle cumulant $v_2\{2\}$.   All the three measurements of $v_2$ are shown. They agree within the systematic errors.
Simulated flows in the NPT scenario are shown in the left panels, 
the lines (top to bottom) correspond to HRG   with $\eta/s$=0,0.08, 0.16, 0.24 and 0.3 respectively. Flows in the PT scenario with QGP viscosity $\eta/s$=0, 0.08, 0.12 and 0.16 are shown in the right panels. For elliptic flow also, we have computed $\chi^2$ values for the fits. The results are shown in Fig.\ref{F3}b. 
In the PT scenario, best fit to the 0-40\% charged particles elliptic flow data is obtained with QGP viscosity $\eta/s$=0.08 ($(\chi^2/N)_{min}$=6.7). 
In the NPT scenario, the best fit is obtained with HRG with viscosity $\eta/s$=0.24, minimum $(\chi^2/N)_{min}$=11.1.  Note that exactly, at these values of viscosity, $p_T$ spectra are also best explained. However, as opposed to the $p_T$ spectra, elliptic flow is better explained with QGP in the initial state rather than with initial HRG.  If both the spectra and elliptic flows are analysed together, minimum $\chi^2$ is similar in both the scenarios,   $\chi^2_{min}\approx$6.1 in the PT scenario and $\approx$7.2 in the NPT scenario (see Fig.\ref{F3}c). It can not be claimed that experimental data are better explained with QGP in the initial state. Quantitatively, both the scenarios give nearly identical description to the data. Hydrodynamical evolution of initial QGP fluid with viscosity to entropy ratio $\eta/s$=0.08,  thermalised in the time scale $\tau_i$=0.6 fm to central energy density $\varepsilon_0$=29.1 $GeV/fm^3$ and that of HRG with $\eta/s$=0.24, thermalised in the time scale $\tau_i$=1 fm to central energy density $\varepsilon_0$=5.1 $GeV/fm^3$ give nearly equivalent description to the PHENIX data for charged particles $p_T$ spectra and elliptic flow in 0-40\% Au+Au collisions. We have not explored all possible initial conditions in PT/NPT scenario. However,
as the minimum $\chi^2/N$ value in NPT/PT scenarios is reasonably small, it is unlikely that with a different initial condition, data could be {\em much} better explained in the PT scenario, as long as HRG at central temperature $T_i$=220 MeV is physical.

One may argue that whether or not QGP is produced do not depend only on the $p_T$ spectra and elliptic flow. Jet quenching, high $p_T$ suppression etc. do give additional support to the claim of QGP formation at RHIC energy collisions.  
 However, high $p_T$ suppression or jet quenching phenomena samples only a restricted phase space, as they are associated with high $p_T$ trigger. Bulk of the matter are at low $p_T$. To claim that QGP is produced in bulk, it is essential that charged particles $p_T$ spectra and elliptic flow are explained in a hydrodynamic model.


   \begin{figure}[t]
\center
 \resizebox{0.40\textwidth}{!}{%
  \includegraphics{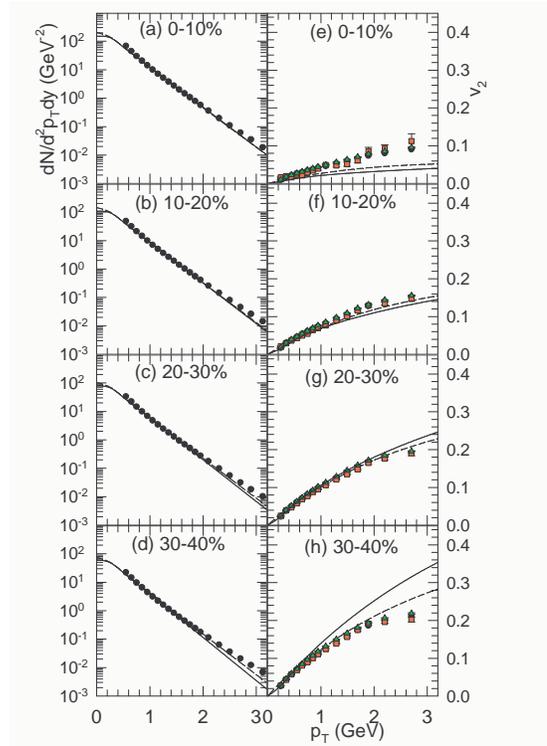}
}
\caption{(color online) In left panels (a)-(d) charged particles $p_T$ spectra in 0-40\% Au+Au collisions are compared with hydrodynamical simulations with
initial HRG and initial QGP. The solid and dashed lines corresponds to evolution of HRG with viscosity to entropy ratio $\eta/s$=0.5, thermalised at
$\tau_i$=3 fm, and QGP with viscosity to entropy ratio $\eta/s$=0.08, thermalised as $\tau_i$=0.6 fm. In panels (e)-(h) results for elliptic flow are shown.}
 \label{F4}
\end{figure}

The present result that if HRG is physical at $T$=220 MeV, NPT and PT scenario   give equivalent description to
RHIC data on $p_T$ spectra and elliptic flow will however be changed if hadronic resonance gas is physical only at a lower temperature $T \leq$ 200 MeV. Even with large viscosity, $\eta/s$=0.5, 
hydrodynamical evolution of HRG, initialised at central temperature $T_i$=200 MeV, do not produce comparable number of hadrons as in experiment, unless
initial or the thermalisation time is large $\tau_i \geq$3 fm. 
In Fig.\ref{F4}, the solid lines are simulation results for charged particles $p_T$ spectra and elliptic flow with viscous ($\eta/s$=0.5) HRG with central temperature $T_i$=200 MeV, thermalised at the time scale $\tau_i$= 3 fm.  For comparison, we have also shown in the simulations results with minimally  viscous ($\eta/s$=0.08)  QGP at  the initial state (the dashed lines). One note that in 0-10\% and 10-20\% collisions, HRG with central temperature $T_i$=200 MeV, thermalised in the time scale $\tau_i$=3 fm, give comparable description to the experimental charged particles $p_T$ spectra, as obtained with QGP in the initial state. In 20-30\% and 30-40\% collisions however, data are better explained with QGP in the initial state. Similarly for the elliptic flow also.  $\chi^2$ analysis also indicate that charged particles $p_T$ spectra and elliptic flow in 0-40\% Au+Au collisions are better expalined with initial QGP than with initial HRG with central temperature $T_i$=200 MeV.
   $\chi^2/N \approx$22  with HRG at the initial state,   factor of $\sim$ 3 larger than that the value obtained with minimally viscous QGP in the initial state.  Apparently, if HRG is physical only at temperature $T\leq$200 MeV, PHENIX data are better explained with initial QGP state than initial HRG state.  
Large thermalisation time $\tau_i$=3 fm, also raises the issue of non-equilibrium contribution to the particle production.
Note that in dissipative hydrodynamics, $\tau_i$ is the time required by the system to achieve near equilibration. Large $\tau_i$ imply that for a substantial time, system remain in a highly non-equilibrium state, which can not be modeled theoretically, and contribution to particle production from this non-equilibrium state, though possibly can not be neglected,  can not be estimated either. The highly non-equilibrium state, with unknown contribution to particle production make the HRG model with limiting temperature $T_i$=200 MeV, untenable. One can possibly exclude initial HRG   
if HRG is physical only at temperature  $T \leq $200 MeV. Though, present lattice simulations indicate that HRG is physical only at $\leq$200 MeV, as discussed earlier, simulations are not precise enough to exclude HRG at $T\approx$220 MeV.
For confirmatory identification of matter produced in RHIC Au+Au collisions with QGP, it is essential to exclude the HRG scenario. Lattice simulations  with realistic boundary conditions, preferably with Wilson fermion actions, are urgently needed to exclude possible HRG scenario.

 


\begin{thebibliography}{99}

\bibitem{BRAHMSwhitepaper}
 BRAHMS Collaboration, I. Arsene {\it et al.},  
Nucl. Phys. A {\bf 757}, 1 (2005). 
 
\bibitem{PHOBOSwhitepaper} 
PHOBOS Collaboration,  B. B. Back {\it et al.},  
Nucl. Phys. A {\bf 757}, 28 (2005). 
 
\bibitem{PHENIXwhitepaper} 
PHENIX Collaboration, K.~Adcox {\it et al.}, 
Nucl. Phys. A {\bf 757} 184 (2005).  
  
\bibitem{STARwhitepaper} 
STAR Collaboration, J. Adams {\it et al.}, 
Nucl. Phys. A {\bf 757} 102 (2005).

  
\bibitem{QGP3}
P.~F. Kolb and U. Heinz,
in {\it Quark-Gluon Plasma 3}, edited by R.~C. Hwa and 
X.-N. Wang (World Scientific, Singapore, 2004), p.~634.



\bibitem{Kolb:2000fha}
  P.~F.~Kolb, P.~Huovinen, U.~W.~Heinz and H.~Heiselberg,
  Phys.\ Lett.\  B {\bf 500}, 232 (2001)
  
\bibitem{Gorenstein:2007mw}
  M.~I.~Gorenstein, M.~Hauer and O.~N.~Moroz,
  Phys.\ Rev.\  C {\bf 77}, 024911 (2008)
  
\bibitem{Demir:2008tr}
  N.~Demir and S.~A.~Bass,
  Phys.\ Rev.\ Lett.\  {\bf 102}, 172302 (2009)
\bibitem{Muronga:2003tb}
  A.~Muronga,
  Phys.\ Rev.\  C {\bf 69}, 044901 (2004)

\bibitem{Pal:2010es}
  S.~Pal,
  Phys.\ Lett.\  B {\bf 684}, 211 (2010)
 
\bibitem{Policastro:2001yc}
  G.~Policastro, D.~T.~Son and A.~O.~Starinets,
  Phys.\ Rev.\ Lett.\  {\bf 87}, 081601 (2001).
  
  
\bibitem{Cheng:2006qk}
  M.~Cheng {\it et al.},
  Phys.\ Rev.\  D {\bf 74}, 054507 (2006)
\bibitem{Cheng:2007jq}
  M.~Cheng {\it et al.},
  Phys.\ Rev.\  D {\bf 77}, 014511 (2008)
  
  
\bibitem{Detar:2007as}
  C.~E.~Detar and R.~Gupta  [HotQCD Collaboration],
  PoS {\bf LAT2007}, 179 (2007)
\bibitem{Karsch:2007dt}
  F.~Karsch,
  PoS {\bf LAT2007}, 015 (2007)
\bibitem{Karsch:2008fe}
  F.~Karsch  [RBC Collaboration and HotQCD Collaboration],
  J.\ Phys.\ G {\bf 35}, 104096 (2008)
  
\bibitem{Cheng:2009zi}
  M.~Cheng {\it et al.},
  Phys.\ Rev.\  D {\bf 81}, 054504 (2010)
  [arXiv:0911.2215 [hep-lat]].


  
\bibitem{Aoki:2006br}
  Y.~Aoki, Z.~Fodor, S.~D.~Katz and K.~K.~Szabo,
  Phys.\ Lett.\  B {\bf 643}, 46 (2006)
  
\bibitem{Aoki:2009sc}
  Y.~Aoki, S.~Borsanyi, S.~Durr, Z.~Fodor, S.~D.~Katz, S.~Krieg and K.~K.~Szabo,
  JHEP {\bf 0906}, 088 (2009)
\bibitem{Fodor:2010zz}
  Z.~Fodor,
  J.\ Phys.\ Conf.\ Ser.\  {\bf 230} (2010) 012013.
  
\bibitem{Borsanyi:2010bp}
  S.~Borsanyi, Z.~Fodor, C.~Hoelbling, S.~D.~Katz, S.~Krieg, C.~Ratti and K.~K.~Szabo
                  [Wuppertal-Budapest Collaboration],
  JHEP {\bf 1009}, 073 (2010)
  [arXiv:1005.3508 [hep-lat]].



\bibitem{Bazavov:2007zz}
  A.~Bazavov and B.~A.~Berg,
  Phys.\ Rev.\  D {\bf 76}, 014502 (2007)

\bibitem{Song:2008si}
  H.~Song and U.~W.~Heinz,
  Phys.\ Rev.\  C {\bf 78}, 024902 (2008).

 
\bibitem{Chaudhuri:2006jd}
  A.~K.~Chaudhuri,
  Phys.\ Rev.\  C {\bf 74}, 044904 (2006)

\bibitem{Chaudhuri:2008ed}
  A.~K.~Chaudhuri,
  J.\ Phys.\ G {\bf 35}, 104015 (2008)

  
  
\bibitem{Chaudhuri:2008sj} A.~K.~Chaudhuri,
 arXiv:0801.3180 [nucl-th].

\bibitem{Chaudhuri:2008je}
  A.~K.~Chaudhuri,
  Phys.\ Lett.\  B {\bf 672}, 126 (2009)
  
\bibitem{Chaudhuri:2009uk}
  A.~K.~Chaudhuri,
  Phys.\ Lett.\  B {\bf 681}, 418 (2009).

\bibitem{Hirano:2009ah}
  T.~Hirano and Y.~Nara,
  Phys.\ Rev.\  C {\bf 79}, 064904 (2009)

\bibitem{Adler:2003au}
  S.~S.~Adler {\it et al.}  [PHENIX Collaboration],
  Phys.\ Rev.\  C {\bf 69}, 034910 (2004)

\bibitem{Afanasiev:2009wq}
  S.~Afanasiev {\it et al.}  [PHENIX Collaboration],
  Phys.\ Rev.\  C {\bf 80}, 024909 (2009)
  
 \end{thebibliography}
\end{document}